\documentclass[aps,prl,twocolumn,superscriptaddress]{revtex4}
\usepackage{graphicx}
\usepackage{latexsym}
\usepackage{amsmath}
\usepackage{lmodern}
\usepackage{color,soul}
\usepackage{ulem}

\newcommand{\BFA}{BaFe$_2$As$_2$}

\newcommand{\FSS}{FeSe$_{1-x}$S$_x$}

\begin{document}

\title{Cryogenic Digital Image Correlation as a Probe of Strain in Iron-Based Superconductors}

\author{Ziye Mo}
\author{Chunyi Li}
\author{Wenting Zhang}
\author{Chang Liu}
\author{Yongxin Sun}
\author{Ruixian Liu}
\author{Xingye Lu}
\email{luxy@bnu.edu.cn}
\affiliation{Center for Advanced Quantum Studies, School of Physics and Astronomy, Beijing Normal University, Beijing 100875, China}

\begin{abstract}
{Uniaxial strain is a powerful tuning parameter that can control symmetry and anisotropic electronic properties in iron-based superconductors. However, accurately characterizing anisotropic strain can be challenging and complex. Here, we utilize a cryogenic optical system equipped with a high-spatial-resolution microscope to characterize surface strains in iron-based superconductors using the digital image correlation method.
Compared with other methods such as high-resolution X-ray diffraction, strain gauge, and capacitive sensor, digital image correlation offers a non-contact, full-field measurement approach, acting as an optical virtual strain gauge that provides high spatial resolution. The results measured on detwinned {\BFA} are quantitatively consistent with the distortion measured by X-ray diffraction and neutron Larmor diffraction.
These findings highlight the potential of cryogenic digital image correlation as an effective and accessible tool for probing the isotropic and anisotropic strains, facilitating the application of uniaxial strain tuning in the study of quantum materials. }
\end{abstract}

\maketitle



In the study of high-temperature superconductivity, understanding the interplay and correlation between various intertwined orders is essential for unveiling the microscopic origins of superconductivity \cite{Keimer2015, Dai2015RMP, Fernandes2022}. In iron-based superconductors (FeSCs), a key issue is the microscopic origin of the electronic nematic phase and its intertwining with superconductivity \cite{Fernandes2022, Fernandes2014, Bohmer2022}.
The electronic nematic phase, defined in the paramagnetic orthorhombic states of FeSCs, is characterized by $C_2$ symmetric electronic anisotropy between the orthorhombic $a$ and $b$ axis. The nematic transition has been shown to coincide with the tetragonal-to-orthorhombic structural transition at $T_s$, driven by electronic instability in a specific electronic degree of freedom (spin, orbital, or charge) (Fig. 1(a)). Above the nematic transition, a nematic fluctuating regime can persist to much higher temperatures (Fig. 1(a)), as evidenced by the widespread nematic susceptibility measured through elastoresistance and other techniques \cite{Chu2010, Chu2012, Kuo2016, Fernandes2014}. In various FeSCs, such as doped {\BFA} and {\FSS}, the nematic transition is gradually suppressed with increasing doping  \cite{Nandi2010, Lu2013, Coldea2021, Liu2024PRL}, leading to the emergence of nematic quantum criticality at the boundary of the nematic phase \cite{Kuo2016, Hosoi2016, Fisher2021}. The electronic nematicity has a significant impact on superconductivity and has been a major focus of research in FeSCs in recent years \cite{Bohmer2022}.

In the study of the electronic nematicity and its correlation with superconductivity, it is often necessary to introduce uniaxial strain as a symmetry-breaking field or as a tuning parameter for the nematicity, as it can directly couple with the nematic order parameter \cite{Bohmer2022}. For instance, nematic susceptibility derived in the mean-field frame is $\chi_{B_{2g}}=d\psi_{B_{2g}}/d\varepsilon_{xy}$ (shear strain $\varepsilon_{xy}\to0$), which is proportional to the elastoresistance coefficient $d\left[(\Delta \rho / \rho)_{x x}-(\Delta \rho / \rho)_{y y}\right]/d\varepsilon_{xx}=c'\cdot
2m_{66}$ (longitudinal uniaxial strain $\varepsilon_{xx}\sim0.01\%$) \cite{Kuo2016}. Uniaxial strain in a larger range $\varepsilon_{xx}\sim0.1\%-1\%$) has been used to tune the critical temperatures of nematic, antiferromagnetic, and superconducting phases \cite{Fisher2021, Ikeda2018, Chu2020, Bartlett2021, Zhao2023}. 

Strain-tuning effects provide valuable insights into the pairing mechanisms and competing interactions of FeSCs \cite{Fernandes2013, Kang2018, Fisher2024}. However, accurately characterizing the uniaxial or anisotropic strain, especially at low temperatures, can be challenging due to the complexity of these materials and the limitations of conventional measurement techniques \cite{Chu2012, Ikeda2018, Sanchez2021, Liu2024PRL}.

\begin{figure}
\includegraphics[width=8.5cm]{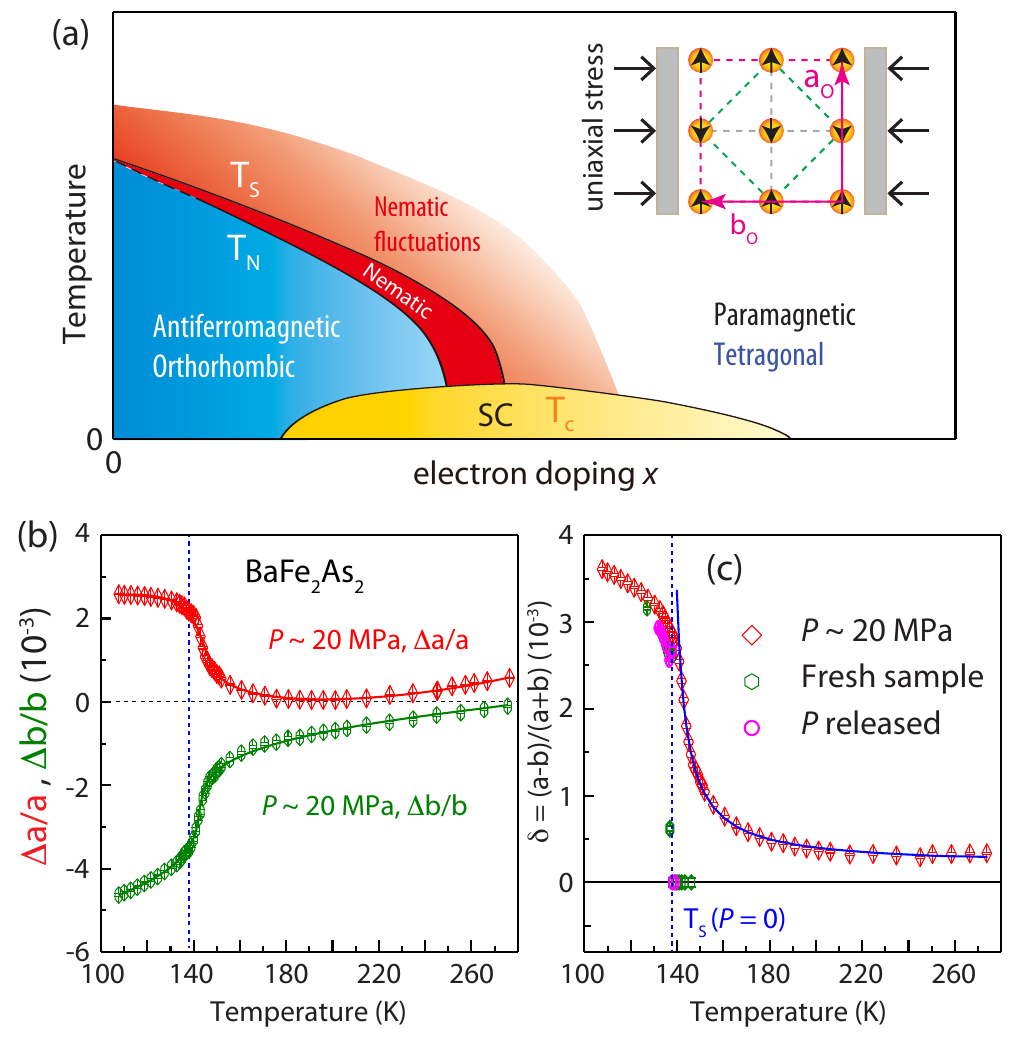}
\caption{(a) Schematic phase diagram of electron-doped {\BFA}. $T_s$, $T_N$ and $T_c$ denotes structural (nematic), magnetic, and superconducting phase transition temperatures, respectively. The antiferromagnetic orthorhombic, nematic, superconducting, and nematic fluctuating regimes are marked by different color regions. The inset shows a schematic of the application of uniaxial-stress on {\BFA} structure. In the inset, the orange filled circles denote Fe$^{2+}$ ions, with their spins denoted by the black arrows. (b) Temperature-dependent relative change of the lattice parameter $a$ and $b$ of {\BFA} under $P\sim20$ MPa uniaxial pressure. (c) Orthorhombic lattice distortion (red diamonds) of the uniaxial-pressured {\BFA} (extracted from (b)) and the unstrained {\BFA} (green and magenta symbols). The blue dashed lines mark the $T_s$ for the unstrained sample. The data in (b) and (c) are reproduced from our neutron Larmor diffraction results reported in ref. \cite{Lu2016}.
}
\label{fig1}
\end{figure}

High-resolution X-ray diffraction (XRD) might be the best method to assess lattice distortions and strain fields in crystalline materials \cite{Sanchez2021, Kim2011}. Synchrotron-based X-ray diffractometer can reach a resolution of $\Delta d/d\sim10^{-4}$ (or $0.01\%$) thanks to the high photon flux of synchrotron beamline \cite{Sanchez2021}. Neutron diffraction---especially the neutron Larmor diffraction technique with super-high resolution in $\Delta d/d\sim10^{-5}-10^{-6}$ --- is an alternative method to measure small strain \cite{Lu2016}. However, these two techniques require access to synchrotron facilities, neutron sources, and sophisticated instrumentation, making them inaccessible to many research laboratories. Lab-based XRD, with much lower photon flux, usually does not have enough resolution to characterize small strain ($\varepsilon\sim0.1\%$). Additionally, XRD and neutron diffraction provide an average strain measurement over a large area of the sample and even the entire sample for neutron diffraction, which may not capture local variations critical to understanding strain effects at the microscopic level.

A simpler and more practical approach for characterizing uniaxial strain over a wide range ($\varepsilon=0-\sim1\%$) is the use of a resistance strain gauge \cite{Chu2012, Kuo2016}. This method is inherently precise and has been widely adopted in previous studies of FeSCs and related materials \cite{Kuo2016, Qian2021}. It is particularly suitable for larger samples, as the typical size of a strain gauge exceeds $1\times1$ mm$^2$, providing an averaged strain measurement similar to that obtained through diffraction techniques. However, the strain gauge must be securely glued to the sample surface, and the strain transmission efficiency from the sample surface to the gauge may be significantly lower than $100\%$ at higher strains \cite{Chu2012}. Additionally, when measuring uniaxial strain at low temperatures, several factors affecting the strain gauge’s resistance must be considered. For example, the resistivity of the strain gauge may show non-linear changes at very low temperatures ($T\lesssim25$ K) (Fig. 2), potentially affecting the accuracy of the strain measurement.

In addition to strain gauges, parallel plate capacitors are also commonly used to determine uniaxial strain by tracking the displacement between the plates. This method allows for accurate strain measurement over a wide range and is the solution employed in commercial strain cells \cite{Bartlett2021}. 
However, in some strain cells, the displacement measured by the capacitor ($\varepsilon_{d}$) arises not only from the strain on the sample ($\varepsilon_{xx}$) but also from the deformation of the epoxy used to secure the crystal. The ratio $\varepsilon_{xx}/\varepsilon_{d}$ has been reported to be approximately 70\% and needs to be characterized using other techniques, such as high-resolution XRD, although it can also be modeled using finite-element analysis \cite{Ikeda2018, Sanchez2021}.

Compared to the aforementioned techniques, the digital image correlation (DIC) method offers an effective alternative for strain measurement, enabling non-contact, full-field analysis of surface displacements and strains with high spatial resolution. 
The fundamental principle of DIC entails capturing a sequence of images of the sample surface under different deformation states or external conditions. By performing a detailed comparison of these images, DIC enables the tracking of the movement of distinct surface features or patterns, thereby facilitating the precise determination of the displacement field across the sample \cite{ncorr, DICe}.

Acting as a virtual strain gauge, DIC is a more accessible and versatile tool for studying anisotropic strain in complex materials such as FeSCs. Although DIC has been used to assess strains in some quantum materials \cite{Sunko2019, Gallo2024}, its accuracy and reliability for cryogenic sample environments require careful validation, a critical aspect that has not yet been fully explored.

In this work, we utilized a custom-built cryogenic digital image correlation (CDIC) system to measure the uniaxial strain of detwinned {\BFA} along the $a/b$ axes (associated with the orthorhombic lattice distortion). The results are quantitatively consistent with those obtained from neutron Larmor diffraction, thereby demonstrating the effectiveness and precision of the cryogenic digital image correlation method. Furthermore, the system’s integration with a piezoelectric driver and a capacitance meter allows for detailed characterization of voltage-dependent strain transmission efficiency across various strain cells. The CDIC technique introduced here proves to be highly effective for the strain characterization and calibration of a wide range of quantum materials, mechanical devices, and piezoelectric strain cells.

To verify the effectiveness of the CDIC method, we selected {\BFA} as a reference, given that its orthorhombic lattice distortion under uniaxial pressure has been previously measured using neutron Larmor diffraction \cite{Lu2016}. As depicted in Figs. 1(b) and 1(c), Larmor diffraction measurements of the orthorhombic (400) and (040) nuclear Bragg peaks of {\BFA} under uniaxial pressure ($P=20$ MPa) from $T=100$ K to 280 K reveal the relative changes in the lattice parameters $a$ and $b$ across the structural transition \cite{Lu2016}. For an unstrained sample ($P = 0$), the lattice parameters are equal ($a = b$) at temperatures above $T_s$. However, with uniaxial pressure ($P = 20$ MPa), the structural transition evolves into a crossover, and the system remains orthorhombic up to $280$ K. Figure 1(c) illustrates the orthorhombic lattice distortion, defined as $\delta=(a-b)/(a+b)$, for both unstrained and strained {\BFA} samples. The resolution $\Delta d/d$ in this experiment is better than $0.001\%$ \cite{Lu2016}.

We first test the effectiveness of resistive strain gauges in measuring anisotropic strain in detwinned {\BFA}. Figure 2(a) illustrates the experimental setup used for the measurements \cite{Lu2014, Lu2022}. A cryogenic strain gauge (CFLA-1-350-11, TML strain gauge) was aligned along the edge and affixed to the surface of a $\sim4\times4\times0.1$ mm$^3$ {\BFA} single crystal, which had been precisely cut from a larger crystal using a high-precision wire saw along the tetragonal $[110]$ and $[1\overline10]$ directions. The crystal was then inserted into a mechanical uniaxial pressure device, with the strain gauge oriented perpendicular to the pressure direction for resistance measurement. This procedure was repeated with the crystal rotated by 90$^\circ$. Pressure was applied using a set of spring washers, activated by tightening a screw at the end of the device.

\begin{figure}
\includegraphics[width=8.5cm]{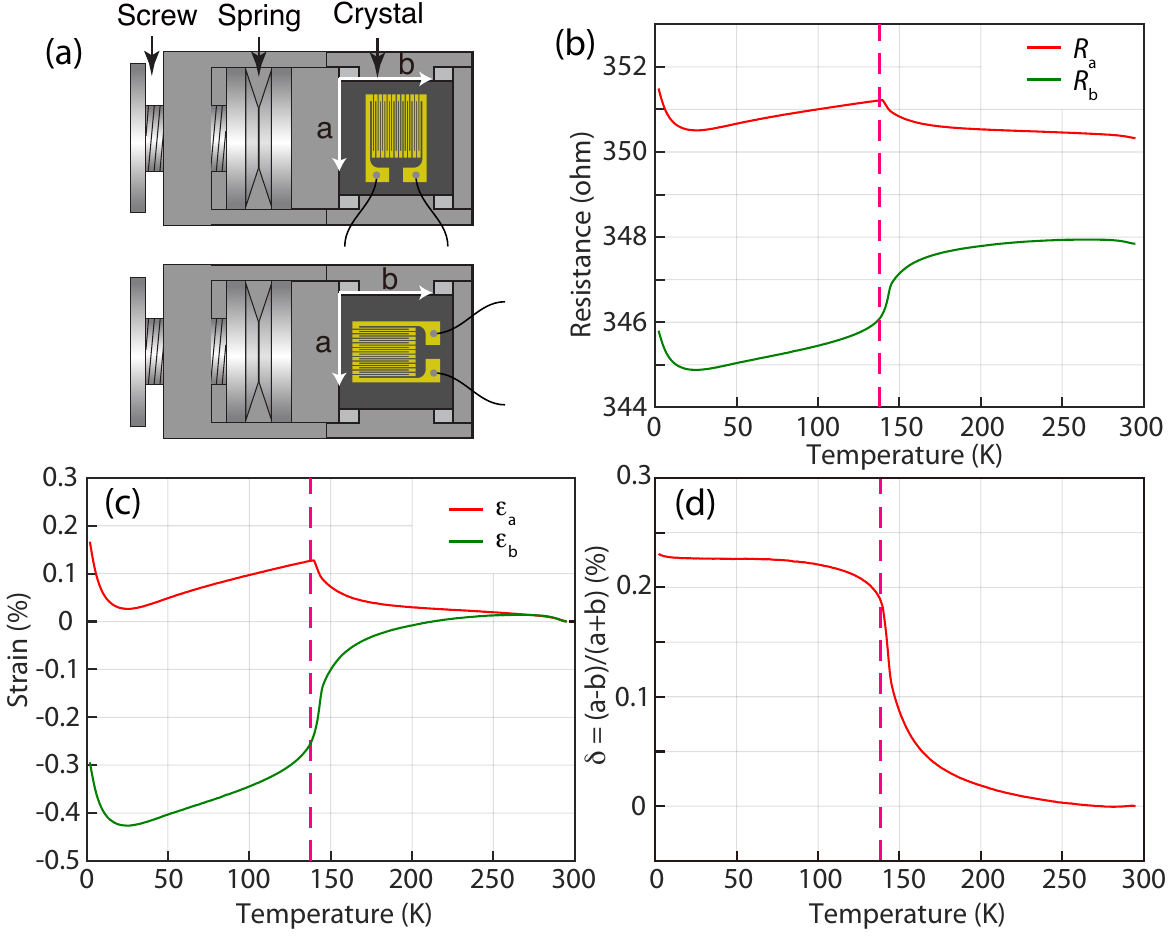}
\caption{(a) Schematic experimental setup for measuring the anisotropic strain (orthorhombic lattice distortion) of detwinned {\BFA} using strain gauge (Part No. CFLA-1-350-11, TML strain gauge). The upper setup measures the resistance of the strain gauge along the $a$ axis ($R_a$) and the lower $b$ axis ($R_b$). (b) Strain-gauge resistance $R_a$ and $R_b$ as a function of temperature. (c) Uniaxial strain along the $a$ axis ($\varepsilon_a$) and the $b$ axis ($\varepsilon_b$), assuming the crystal is unstrained ($\varepsilon=0$) at $T=295$ K and the resistance-change is purely induced by uniaxial strain. (d) Orthorhombic lattice distortion $\delta$ [$2\delta=(\varepsilon_a-\varepsilon_b)$] extracted from $\varepsilon_a$ and $\varepsilon_b$ as shown in (c).}
\label{fig2}
\end{figure}

Figure 2(b) presents the raw data for the strain-gauge resistance measured along the $a$ and $b$ axes, denoted as $R_a$ and $R_b$, respectively. The resistance is sensitive to strain along the gauge length, with the conversion relation given by $K\varepsilon = \frac{\Delta R}{R_0}$, where $K\approx2$ is the gauge factor. The temperature dependence of the resistance roughly aligns with the data shown in Figure 1. The observed difference at $T=295$ K between $R_a$ and $R_b$ may result from changes in the experimental setup rather than being induced by uniaxial pressure. The upturns at low temperatures ($T\lesssim25$ K) could be attributed to changes in the resistivity of the strain gauge. Assuming $R_a = R_b$ at $T=295$ K and that the temperature-dependent resistivity change is induced by uniaxial strain, we converted the resistance data to uniaxial strain, as shown in Fig. 2(c), and subsequently obtained the orthorhombic lattice distortion in Fig. 2(d). The strains $\varepsilon_a(T)$ and $\varepsilon_b(T)$ differ from the actual strains (Fig. 1) because several factors influencing resistance, such as the anomalous upturn at low temperatures, apparent strain, and thermal output of the strain gauge, were not subtracted. Interestingly, these factors appear to cancel out when calculating $\delta$ from $\varepsilon_a(T) - \varepsilon_b(T)$. The resulting $\delta(T)$ exhibits the same temperature-dependent behavior as shown in Figure 1(c), with the maximum value of $\delta\approx0.23\%$, which is approximately $70\%$ of the actual value. This finding is consistent with previous reports indicating that strain transmission efficiency can be significantly lower than $100\%$ \cite{Ikeda2018, Chu2020, Sanchez2021}.

\begin{figure}
\includegraphics[width=8cm]{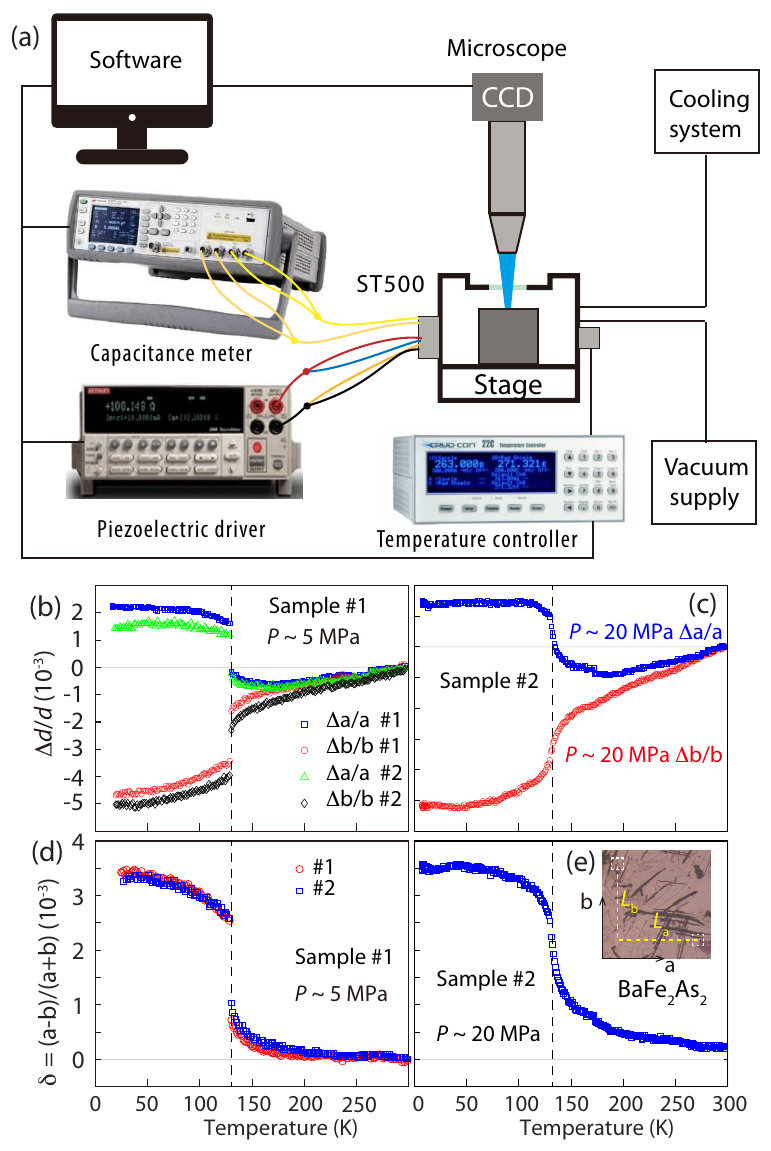}
\caption{(a) Cryogenic uniaxial-strain-tuning system equipped with a high-spatial-resolution microscope used to characterize the surface strain of a single crystal with the CDIC method. The essential part of this system is a liquid helium cryostat (ST500) with an optical window. The cryostat was placed onto a five-axis stage---translations along the X, Y, and Z directions, rotation ($\phi$) and tilting $\chi$ along one edge. A piezoelectric stack driver and a capacitance meter can be connected to a strain cell (installed in the cryostat) via coaxial feedthrough cables. (b), (c) Relative change of lattice parameter $\Delta a/a$ and $\Delta b/b$ measured on uniaxially-pressured {\BFA} with $P\sim5$ MPa on sample \#1 (b) and $P\sim20$ MPa on sample \#2 (c). (d), (e) Orthorhombic lattice distortions extracted from the results in (b), (c). The inset in (e) shows the cleaved surface of a {\BFA} crystal ($1\times1$mm$^2$) for measurements. The scratches are intentionally made to facilitate the tracking of specific positions.}
\label{fig3}
\end{figure}

Figure 3(a) illustrates a schematic diagram of our CDIC system, which is also a cryogenic uniaxial-strain-tuning system suitable for both transport and optical studies. The system features a liquid helium flow cryostat (ST500) equipped with an optical window, which enables clear imaging of the sample surface at cryogenic temperatures. A digital metallographic microscope is used to capture high-resolution images of the sample surface, allowing for precise tracking of surface feature displacements. Additionally, the system is outfitted with a capacitance meter and a piezoelectric driver, making it compatible with piezoelectric strain cells. All meters are connected to a computer via GPIB cables. The meters and microscope can be controlled using LabVIEW-based software, ensuring streamlined operation and data acquisition.

We employed the same mechanical uniaxial pressure device depicted in Fig. 2(a) to detwin a cleaved {\BFA} crystal. To facilitate DIC analysis, multiple scratches were made on the crystal surface to enhance image contrast and ensure the unique identification of surface features (inset of Fig. 3(e)). A series of images were captured while warming the sample from approximately $10–20$ K to $300$ K. The inset in Fig. 3(e) shows a portion of one such image for the {\BFA} crystal. Several DIC software packages, such as the DIC Engine (DICe) \cite{DICe} and Ncorr \cite{ncorr}, can be utilized to analyze the acquired images and calculate the displacement of each feature point \cite{DICe}. These software tools typically employ correlation algorithms to compare subsets of pixels between the reference and deformed images. From the resulting displacement field, strain can be computed as the spatial gradient of displacement, providing a detailed map of strain distribution across the sample.
 
To measure the $\varepsilon_a$ and $\varepsilon_b$ in {\BFA}, we adopted a simpler approach by tracking the relative displacement of two feature regions aligned along the diagonal directions (highlighted by white dashed squares in Fig. 3(e)). The distances between the centers of these regions, denoted as $L_a$ along the $a$ axis and $L_b$ along the $b$ axis, were used to calculate the strain as $\varepsilon_{a/b} = \left(\frac{\Delta L}{L}\right)_{a/b}$.

For data analysis, we utilized a set of MATLAB-based DIC codes to compare the deformed image with the reference image. The position of each feature region in the new image was determined by calculating the sum of the squared differences in grayscale values between the scanned and reference regions. Specifically, the variance was calculated as $V(x, y) = \sum_{i,j}(G_{i,j} - G^{\prime}_{i+x, j+y})^2$, with the location corresponding to the minimum variance identifying the new position. To enhance spatial resolution, the algorithm incorporated linear interpolation of the grayscale matrix.

Figures 3(b) and 3(c) show the $\varepsilon_{a}(T)=\Delta a(T)/a(T=300$K) and $\varepsilon_{b}(T)=\Delta b(T)/b(T=300$K) calculated from two groups of feature regions on sample $\#1$ under $P\sim5$ MPa and one group of feature regions on sample $\#2$ under $P\sim20$ MPa, respectively. Figures 3(d) and 3(e) depict the orthorhombic distortion calculated from $\varepsilon_{a/b}$ shown in Figs. 3(b) and 3(c). Both the $\varepsilon_{a/b}(T)$ and the $\delta(T)$ are quantitatively consistent with the Larmor diffraction results in Fig. 1, except that the offset between $a$ and $b$ at room temperature cannot be determined directly in our case.

\begin{figure}
\includegraphics[width=8.5cm]{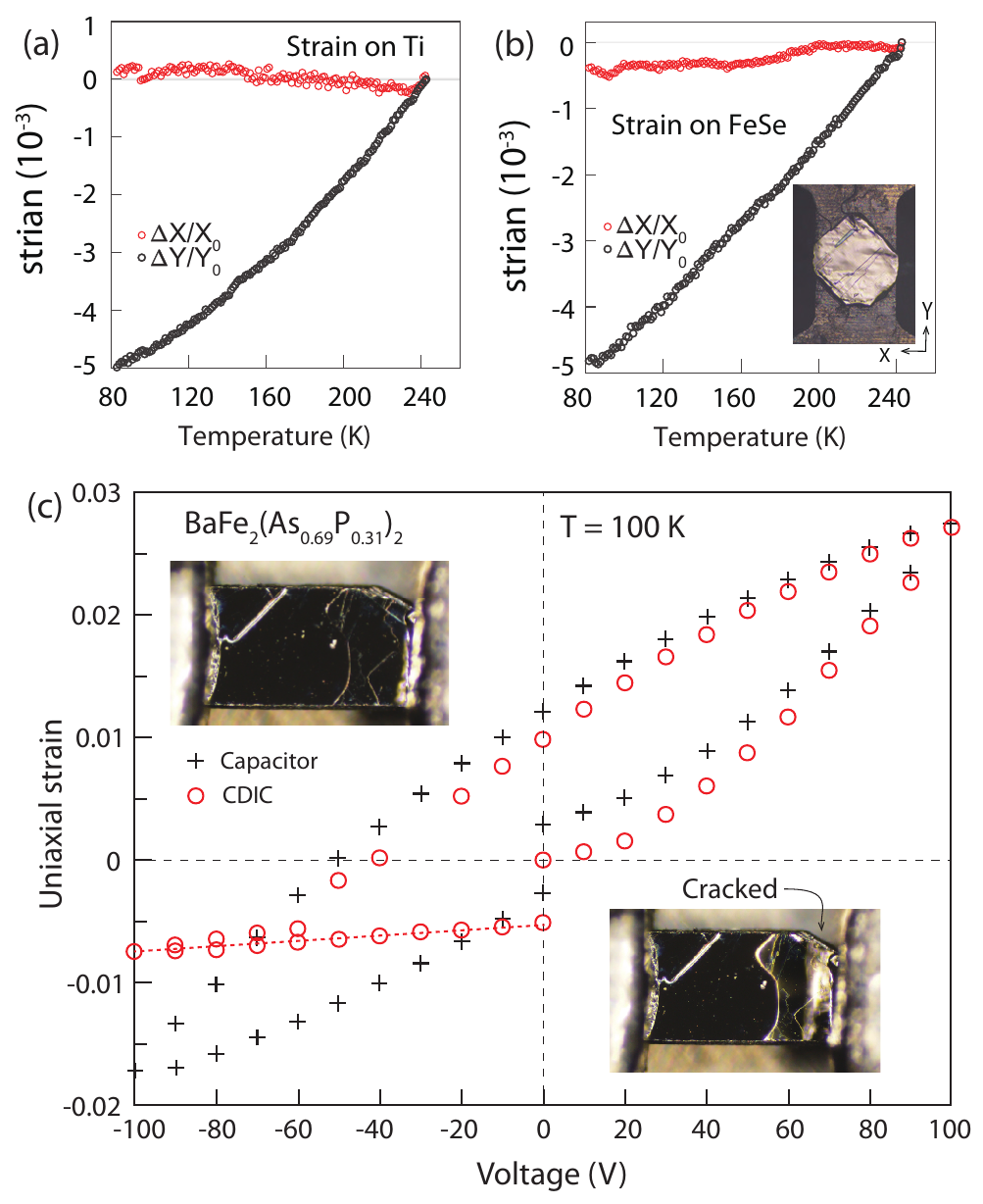}
\caption{(a), (b) CDIC measurement of the anisotropic strain of a titanium platform (a) and the FeSe single crystal (b) glued on the platform (the inset in (b)). In this setup, the platform is inserted into a mechanical strain device based on differential thermal expansions of invar alloy, titanium, and aluminum alloy. While cooling, the uniaixal strain applied onto the platform can be transferred to the FeSe thin crystal. (c) Uniaxial strain characterization of a BaFe$_2$(As$_{0.69}$P$_{0.31}$)$_2$ ($0.25\times0.08\times0.4$ mm$^3$) single crystal installed on a commercial strain cell driven by piezoelectric stacks. The black crosses and red circles represent the strain values collected by capacitance sensor in the strain cell and the DIC method. The left upper inset and the right lower inset are the photos of the sample before and after the measurements, respectively.}
\label{fig4}
\end{figure}

At $P\sim5$ MPa, although the structural transition become a crossover, the dramatic change of $\delta$ at $T_N$ remains (Figs. 3(b) and 3(d)) \cite{Chu2010, Kim2011}. In comparison, the $\delta(T)$ become a smooth crossover across the $T_N$ and $T_s$, which are consistent with previous X-ray and neutron diffraction results on {\BFA} \cite{Lu2016}.
In Fig. 3(e), we make zero-point compensation $\delta(T=300$K$)=0.02\%$ to account for the distortion generated by the pressure under $P\sim20$ MPa. At low temperatures, $\delta$ reaches $0.35\%$, which is also very close to the value ($\delta=0.36\%$) measured by high-resolution diffraction methods. 
The spatial resolution $\Delta d/d$ estimated from Fig. 3(b) is $\sim0.01\%$ (Fig. 3).
These results conclusively demonstrate the effectiveness and efficiency of the CDIC method for characterizing strains in FeSCs. This method is also compatible with custom-built strain devices and piezoelectric strain cells.

In Figs. 4(a) and 4(b), we illustrate the strain characterization of a custom-built mechanical strain device, which operates based on the differential thermal expansion coefficients of invar alloy, titanium, and aluminum. This device was used to generate an anisotropic strain $(\varepsilon_{xx}$ - $\varepsilon_{yy})$ ranging from approximately $-0.4\%$ to $-0.8\%$ in our previous study of nematic spin correlations in {\FSS} \cite{Liu2024PRL}. The device applies uniaxial strain to a titanium platform (bridge) through the differential thermal contraction between the aluminum frame and the inner invar-alloy blocks that hold the titanium platform. A FeSe single crystal was affixed to this platform via a thin layer of epoxy, allowing the uniaxial strain applied to the platform to be transmitted to the crystal.
Fig. 4(a) shows the anisotropic strain on the titanium bridge, which can be accurately measured with DIC, reaches $(\varepsilon_{xx}$ - $\varepsilon_{yy}) \approx -0.5\%$ when cooled from 240 K to 80 K. Simultaneously, the strain on the FeSe crystal (Fig. 4(b)) was also measured with high precision, revealing a transmission efficiency of approximately $90\%$. The modified ST500 cryostat offers ample space to accommodate a variety of custom-built strain devices \cite{Liu2024PRL, Tanatar2016}.

For the piezoelectric strain cell, Fig. 4(c) shows the uniaxial strain of a BaFe$_2$(As$_{1-x}$P$_x$)$_2$ single crystal (mounted on a Razorbill FC100 strain cell) measured simultaneously by a capacitive sensor (black crosses) and the CDIC method (red circles) at $T = 100$ K \cite{Zhao2023}. The maximum strain reached 3\%. The voltage driving the strain cell was varied from 0 to 100 V, 100 V to -100 V, and -100 V back to 0 V.

During the first part of the loop ($0\to100\to-60$), the strain calculated from the capacitance closely follows the actual strain measured by CDIC. However, as the voltage decreases beyond -60 V, a deviation emerges. While the strain measured by the capacitive sensor completes the loop, the strain measured by DIC does not change during the rest of the loop. This discrepancy occurs because the crystal was cracked at one end (the inset of Fig. 4(c)), preventing effective displacement transmission to the crystal. The CDIC strain value at zero voltage corresponds to the true zero-strain position. According to these results,  we find the strain measured by CDIC is more representative of the true strain.

In summary, using {\BFA} as a reference, we establish the cryogenic digital image correlation (CDIC) system and method as a powerful probe of the strains on FeSCs.  
This approach overcomes the limitations of traditional strain measurement techniques by offering a non-contact, high-resolution ($\Delta d/d\sim10^{-4}$) alternative capable of capturing the temperature-dependent anisotropic strain in complex materials and devices \cite{Sunko2019, Tanatar2016, Liu2024PRL, Yang2023}. The versatility of this approach suggests that it can be extended to a wide range of quantum materials, paving the way for future research in strain-engineered quantum materials.

The work is supported by the National Key Projects for Research and Development of China (Grant No. 2021YFA1400400), the Fundamental Research Funds for the Central Universities, and the National Natural Science Foundation of China (Grant Nos. 12174029 and 11922402).

\end{document}